\documentclass[prl,twocolumn,floatfix,superscriptaddress]{revtex4-1}
\usepackage{dcolumn,amsmath}
\usepackage[dvips]{graphicx}
\usepackage{bm}
\usepackage{hyperref}

\newcommand{\Eeff}{\ensuremath{E_{\rm eff}}}
\newcommand{\eEDM}{{\em e}EDM}

\begin{document}
\title{Towards the search of T,P-odd interactions in lead monofluoride, PbF}
\author{L. V.\ Skripnikov}\email{leonidos239@gmail.com}
\author{A. D.\ Kudashov}
\author{A. N.\ Petrov}
\author{A. V.\ Titov}
\homepage{http://www.qchem.pnpi.spb.ru}
\affiliation{B.P.Konstantinov Petersburg Nuclear Physics Institute, Gatchina, Leningrad district 188300, Russia}
\affiliation{Dept.\ of Physics, Saint Petersburg State University, Saint Petersburg, Petrodvoretz 198504, Russia}

\date{}
\begin{abstract}
The relativistic coupled-clusters method combined with the generalized relativistic effective core potential approach and nonvariational one-center restoration technique is applied to evaluation of parameters of spin-rotational effective Hamiltonian in lead monofluoride to study the effects of violation of time-reversal invariance (T) and space parity (P) in PbF. The obtained hyperfine structure 
constants, A$_{||}$=9942 MHz and A$_{\perp}$=-7174 MHz are stable with respect to the improvement of the correlation treatment
and they are in a very good agreement with the experimental data, A$_{||}$=10147 MHz and A$_{\perp}$=-7264 MHz [PRA {\bf 84}, 022508 (2011)]. This is essential to the important task of verifying the value of effective electric field \Eeff =40 GV/cm,
the parameter of P-odd interaction $W_{P}$=-1213~Hz and the parameter of T,P-odd pseudoscalar$-$scalar electron$-$nucleus interaction $W_{T,P}$=91~kHz, which are of primary interest in the paper.

%
\end{abstract}

\maketitle


\section{Introduction}

Theoretical study of PbF is primarily motivated by the proposed experiments to search for the effects of simultaneous violation of time-reversal invariance (T) and space parity (P) (T,P-parity nonconservation or PNC effects below) such as electron electric dipole moment (\eEDM) \cite{Kozlov:87, Dmitriev:92, Shafer-Ray:06, Shafer-Ray:08E, Baklanov:10, Petrov:13}.
   Besides, an experiment using PbF to measure the P-odd anapole moment has also been recently suggested \cite{Alphei:11}. The hyperfine structure (HFS) constants for the ground state of PbF are known with high accuracy \cite{Mawhorter:11, Petrov:13}.

First, the PbF molecule
was
theoretically studied
in Refs.~\cite{Kozlov:87, Dmitriev:92} in order to evaluate the parameters of P- and T,P-odd effects in heavy-atom molecules. Then a number of new studies was performed in Refs.~\cite{Meyer:08, Baklanov:10,Borschevsky:13}. However, even the most recent results \cite{Baklanov:10,Borschevsky:13} display a significant disagreement for the parameter of P-odd interaction (see Eq.~(\ref{W_p}) below). It can be considered rather surprising, especially, if one takes into account that PbF, unlike other systems actively considered for PNC experiments to date (HfF$^+$ \cite{Cossel:12}, YbF \cite{Hudson:11a}, ThO \cite{ThO}, ThF$^+$ \cite{Cornell:13}, WC \cite{Lee:13a} etc.), is not a compound of a transition $d$-element or lanthanide/actinide $f$-element. Nevertheless, the main feature of the electronic structure of PbF $\Pi_{1/2}$ ground state is that the valence unpaired electron is occupying the $\pi-$state (which is derived from the $6p-$state of Pb), as opposed to the majority of other systems, taken into consideration over the years, having unpaired $\sigma-$electron (that is derived from the valence $s-$state of a heavy atom). These unpaired electrons provide the leading contribution to the HFS constant
\footnote{
The hyperfine structure constants are traditionally used as
  ``benchmark'' characteristics for
checking the accuracy of the studied molecular parameters ``linking'' the fundamental PNC phenomena like electron electric dipole moments, anapole and Schiff moments, etc.\ with the induced molecular PNC effects see \cite{Kozlov:95,Titov:06amin,Titov:14a}.}
and PNC parameters of interest. However, the leading (one-configuration) contribution is somewhat suppressed for the systems with unpaired $\pi-$state(s) due to their 
   weaker
 asymptotic behaviour on the heavy nucleus, making the relative influence of correlation effects, etc.\
more important. This means that the PbF results should be
  sensitive to the quality of calculation,
i.e., accounting for correlation effects and 
their 
interplay
with spin-orbit interaction.

A revival of interest in PbF during last years \cite{Alphei:11} was initiated by the discovery of coincidental near degeneracy for levels of opposite parity in the ground rotational state J = 1/2 for $^{207}$PbF \cite{Shafer-Ray:08}, caused by the near cancellation between the shifts in the energies of these levels due to omega-type doubling and the magnetic hyperfine interaction.
This degeneracy had contradicted the previous theoretical studies and was resolved in \cite{Baklanov:10}. This leads to improved sensitivity of PbF to the PNC effects, as it was earlier expected. Besides, the molecule was suggested to be sensitive to variation of the fundamental constants, \cite{Flambaum:2013}. So, a reliable theoretical study of the parameters of the P- and T,P-odd effective spin-rotational Hamiltonian is of considerable interest for experiment.

\section{Theory}

Following Refs. \cite{Kozlov:87, Dmitriev:92}, we represent the effective spin-rotational Hamiltonian as
\begin{gather}
\mathbf{H}_{\rm eff} = 
B\mathbf{J}^2 + \Delta\mathbf{J}\cdot\mathbf{S}^{\prime} + \mathbf{I}\cdot\widehat{A} \cdot \mathbf{S}^{\prime}+ \nonumber\\
\mu_{B}~\mathbf{B}\cdot\widehat{G}\cdot
\mathbf{S}^{\prime}+D~\mathbf{E}\cdot\mathbf{n}\nonumber\\
(W_{p}~\kappa_{p})~\mathbf{n\times S}^{\prime}\cdot
\mathbf{I}\mathbf{+(}W_{T,P}~\kappa_{T,P}+W_{d}~d_{e})\mathbf{S}^{\prime}\cdot\mathbf{n}
\label{HEFF}
\end{gather}
The first line in eq. (\ref{HEFF}) corresponds to the rotational and hyperfine structure.
The second line describes the interaction with the external magnetic ($\mathbf{B}$) and electric ($\mathbf{E}$) fields.
The last line corresponds to $P-$ and $T,P-$odd interactions.
$B$ is the rotational constant, $\Delta$ is the $\Omega$-doubling constant, $\mathbf{n}$
is the unit vector directed from the heavy nucleus to the light one, $\mathbf{S}^{\prime}$ is
effective spin \cite{Kozlov:87, Dmitriev:92}.
In the molecular frame, the tensor contractions
\begin{gather}
\mathbf{I}\cdot\widehat{A}\cdot \mathbf{S}^{\prime}=A_{||}\mathbf{I}_{0}\mathbf{S}_{0}^{\prime}-A_{\perp}(\mathbf{I}_{1}\mathbf{S}_{-1}^{\prime}
+\mathbf{I}_{-1}\mathbf{S}_{1}^{\prime}),
%
\end{gather}
are determined by the hyperfine parameters $A_{||},$ $A_{\perp}$.
%
Equations for the parameters of the effective spin-rotational Hamiltonian (\ref{HEFF}) are given below.

To obtain the $W_d$ parameter one can evaluate the expectation value of the following T,P-odd operator (discussed in Refs.~\cite{Kozlov:87, Kozlov:95, Titov:06amin}):
\begin{equation}
\label{matrelem}
W_d = \frac{1}{\Omega}
\langle \Psi_{^2\Pi_{\pm 1/2}}|\sum_i\frac{H_d(i)}{d_e}|\Psi_{^2\Pi_{\pm 1/2}}\rangle,
\end{equation}
where $d_e$ is the value of \eEDM, $\Psi$ is the wave function of the considered state of PbF molecule,
$\Omega= \langle\Psi_{^2\Pi_{\pm 1/2}}|\bm{J}\cdot\bm{n}|\Psi_{^2\Pi_{\pm 1/2}}\rangle$,
$\bm{J}$ is the total electronic momentum, $\bm{n}$ is the unit vector along the molecular axis directed from Pb to F
($\Omega{=} \pm 1/2$ for the considered $^2\Pi_{\pm 1/2}$
state of PbF),
\begin{eqnarray}
  H_d=2d_e
  \left(\begin{array}{cc}
  0 & 0 \\
  0 & \bm{\sigma E} \\
  \end{array}\right)\ ,
 \label{Wd}
\end{eqnarray}
$\bm{E}$ is the inner molecular electric field, and $\bm{\sigma}$ are the Pauli matrices. In these designations a widely used parameter known as the effective electric field acting of unpaired electrons is $E_{\rm eff}=W_d|\Omega|$.

The T,P-odd pseudoscalar$-$scalar electron$-$nucleus interaction with a characteristic dimensionless constant $k_{T,P}$ is given by the following operator (see \cite{Hunter:91}):
\begin{eqnarray}
  H_{T,P}=i\frac{G_F}{\sqrt{2}}Zk_{T,P}\gamma_0\gamma_5n(\textbf{r}),
 \label{Htp}
\end{eqnarray}
where $G_F$ is the Fermi-coupling constant, $\gamma_0$ and $\gamma_5$ are the Dirac 
matrices and $n(\textbf{r})$ is the nuclear density normalized to unity. To extract the fundamental $k_{T,P}$ constant from an experiment one needs to know the factor $W_{T,P}$ that is determined by the electronic structure of a studied molecular state on a given nucleus:
\begin{equation}
\label{WTP}
W_{T,P} = \frac{1}{\Omega}
\langle \Psi_{^2\Pi_{\pm 1/2}}|\sum_i\frac{H_{T,P}(i)}{k_{T,P}}|\Psi_{^2\Pi_{\pm 1/2}}
\rangle\ .
\end{equation}

The $P-$odd electron$-$nucleus interaction is characterised by the dimensionless constant $k_{P}$, which is given by
\begin{equation}
 H_{P} = k_{P}\frac{G_\mathrm{F}}{\sqrt{2}}
\bm{\alpha}\cdot\bm{I}
n(\textbf{r}).
\label{HP}
\end{equation}
To extract the fundamental $k_{P}$ constant from an experiment one needs to know the electronic parameter $W_{P}$, which can be written as
\begin{equation}
     W_{p}=\frac{G_\mathrm{F}}{\sqrt{2}}
     \left\langle \Psi_{^{2}\Pi_{1/2}} \left\vert
     n(\textbf{r}) {\alpha_+}
     \right\vert \Psi_{^{2}\Pi_{-1/2}} \right\rangle,
\label{W_p}
\end{equation} 
where $\alpha_+$ is defined as
\begin{eqnarray*}
  \alpha_+=\alpha_x+\mathrm{i}\alpha_y=
  \left(\begin{array}{cc}
  0      & \sigma_x \\
  \sigma_x & 0 \\
  \end{array}\right)+
  \mathrm{i}\left(\begin{array}{cc}
  0      & \sigma_y \\
  \sigma_y & 0 \\
  \end{array}\right).
\end{eqnarray*}

To compute $A_{||}$ and $A_{\perp}$ on Pb in the ground electronic $^2\Pi_{1/2}$ state of PbF molecule the following matrix element can be evaluated:
\begin{eqnarray}
\label{Apar}
A_{\parallel} = 
  \frac{\mu_{\rm Pb}}{I\Omega}
   \langle
   \Psi_{^2\Pi_{\pm 1/2}}|\sum_i\left(\frac{\bm{\alpha}_i
\times
\bm{r}_i}{r_i^3}\right)
_z|\Psi_{^2\Pi_{\pm 1/2}}\rangle,
\end{eqnarray}
\begin{eqnarray}
\label{Aperp}
A_{\perp} = 
  \frac{\mu_{\rm Pb}}
{I}
   \langle
   \Psi_{^2\Pi_{1/2}}|\sum_i\left(\frac{\bm{\alpha}_i
\times
\bm{r}_i}{r_i^3}\right)
_+|\Psi_{^2\Pi_{-1/2}}\rangle  ,
\end{eqnarray}
where $\mu_{\rm Pb}$ is the nuclear magnetic moment of a Pb isotope
with spin $I$,
$ \bm{\alpha}=
  \left(\begin{array}{cc}
  0 & \bm{\sigma} \\
  \bm{\sigma} & 0 \\
  \end{array}\right).
$


\section{Results and discussions}

The matrix elements (\ref{matrelem},\ref{WTP},\ref{W_p},\ref{Apar},\ref{Aperp}) are examples of the so-called ``atom in a compound'' or AIC properties \cite{Titov:14a}.
They are mean values of operators heavily concentrated in the atomic core of Pb and are sensitive to variation of core-region spin densities of the valence electrons. The matrix elements were computed using a scheme, which combines the generalized relativistic effective core potential (GRECP) approach \cite{Mosyagin:10a,Titov:99} with the non-variational restoration procedure \cite{Titov:06amin}. Compared to direct all-electron Dirac-Coulomb four-component methods, the GRECP approach has an advantage of naturally combining the fully-relativistic and scalar-relativistic treatment. Within the latter it is possible in many cases to take account of high-order correlation effects and effects of extended basis set. Moreover, Breit interaction (as well as quantum electrodynamic and other important effects) can be effectively included into the GRECP operator \cite{Petrov:04b, Mosyagin:06amin, Mosyagin:10a}. Although the GRECP approximation and the nonvariational restoration procedures introduce a certain theoretical uncertainty, contemporary full-electron studies have not yet been able to unambiguously surpass our approach when it comes to AIC and spectroscopic properties of interest, as one can see from the recent comparative study of ThO \cite{Skripnikov:14b}.

The single-reference two-component relativistic coupled-clusters method with single, double and perturbative treatment of triple cluster amplitudes, 2c-CCSD(T), was used to take account of both the relativistic and correlation effects for valence electrons. For calculation of off-diagonal matrix elements (\ref{W_p},\ref{Aperp}) the linear-response two-component coupled clusters method with single and double cluster amplitudes \cite{Kallay:5} was used. The $1s-4f$ inner-core electrons of Pb were excluded from the molecular correlation calculations using the ``valence'' semi-local version of the GRECP operator \cite{Mosyagin:10a}. Thus, 31 electrons ($5s^2 5p^6 5d^{10} 6s^2 6p^2$ (Pb) and $1s^2 2s^2 2p^5$(F)) were treated explicitly in our correlation calculations. 
A basis set consisting of $13s$, $12p$, $8d$, $3f$ and $1g$ ([13,12,8,3,1])  contracted Gaussian basis functions on Pb and $6s$, $5p$, $4d$, $3f$ ([6,5,4,3]) functions on F was used. The basis set on Pb was re-optimized from a previous paper \cite{Isaev:00}, while for F the aug-ccpVQZ basis set \cite{Kendall:92} with removed two g-type basis functions was employed. In the present calculation we used Pb---F internuclear distance of 3.9 a.u., which is close to the experimental datum 3.89 a.u.~\cite{Huber:79}.
The coupled-clusters calculations were performed using the {\sc dirac12} \cite{DIRAC12} and {\sc mrcc} \cite{MRCC2013} codes. The nonvariational restoration code developed in \cite{Skripnikov:13b, Skripnikov:13c, Skripnikov:11a} and interfaced to these codes was used to restore the four-component electronic structure near the Pb nucleus. 

Results of calculations, as well as the results of previous studies are given in table \ref{TResults}.
According to our analysis the main cause of difference between values of \Eeff\ obtained in \cite{Baklanov:10} and in the present paper is the contribution of outer-core electrons that was not    considered in \cite{Baklanov:10} within 13-electron calculations.

Taking into account the results from table \ref{TResults} and our earlier studies within the two-step procedure and the coupled clusters approach (e.g., see \cite{Skripnikov:14b,Kudashov:14}) of \Eeff, $W_{T,P}$ and $A_{||}$ we expect that the theoretical uncertainty for our final values is smaller than 7\%.

\begin{table*}[!h]
\caption{Calculated values of molecule-frame dipole moment ($d$), effective electric field (\Eeff), the parameter of the T,P-odd pseudoscalar$-$scalar electron$-$nucleus neutral currents interaction ($W_{T,P}$),
the hyperfine structure constants (A$_{||}$, A$_{\perp}$) and the parameter of $P-$odd electron$-$nucleus interaction ($W_{P}$) 
of the $^2\Pi$ state of PbF.
}
\label{TResults}
\begin{tabular}{ l  c  c  c  c  c  c }
\hline\hline
 Method                             &   $d$,        & \Eeff, & $W_{T,P}$, & A$_{||}$   & A$_{\perp}$, & $W_P$, 
\\
                                    &   Debye       & GV/cm  & kHz        & MHz& MHz & Hz 
\\
\hline      
SCF $^a$
, \cite{Kozlov:87} &    6.1       & 29  $\pm$ 8    & 75  $\pm$ 24         & 9120 $\pm$ 430     & -7850 $\pm$ 390 &  950  $\pm$ 300  
\\ 
SCF $^a$
, \cite{Dmitriev:92}  &   4.62       &  29    &   55       & 10990    & -8990      &    -720
 \\ 

  Semiempirical, \cite{Meyer:08} &          & 31     &           &       &   &        \\
  13e-SODCI$^b$, \cite{Baklanov:10} &   4.26        & 33     & 75         & 9727     & -6860 & -990 
\\ 
  13e-SODCI + OC correction$^c$, \cite{Baklanov:12m} &   5.00        & 37     & 83         & 10262     & --- & --- \\
  DHF+CP$^d$, \cite{Borschevsky:13}                       &  ---          & ---    & ---      &  ---  &---& -1269
\\    

\\
  31e-2c-CCSD                       &   3.97        & 41     & 93     & 10265     & -7174&-1213 
\\ 
  31e-2c-CCSD(T)                    &   3.87     & 40     &  91    & 9942      & ---  &--- 
  \\
  ~~~~~~(Final)                   &        &       &      &        &     \\  
Experiment, \cite{Mawhorter:11}     &   3.5 $\pm$0.3&   ---  &   ---      & 10147  & -7264 & ---       \\  
 
\hline\hline
\end{tabular}
\\
$^a$ SCF, self consistent field.
\\
$^b$ 13-electron SODCI, spin-orbit direct configuration interaction, \cite{Baklanov:10}.
\\
$^c$ 13-electron SODCI plus correction on correlation of the outer-core electrons. The calculation was 
aimed at AIC properties.
\\
$^d$ DHF+CP, Dirac-Hartree-Fock plus core polarization, \cite{Borschevsky:13}. 
\end{table*}

\section{Conclusions}

The relativistic coupled-clusters method combined with the generalized relativistic effective core potential approach and nonvariational one-center restoration technique is applied to evaluate   parameters of spin-rotational effective Hamiltonian in lead 
monofluoride, which can be used to study the effects of T and P symmetries' violation in molecular beam or Stark trap \cite{Alphei:11} experiments.
Our results are rather stable with respect to the improvement of the correlation treatment. Besides, hyperfine structure constants (A$_{||}$, A$_{\perp}$), and molecule-frame dipole moment are in a very good agreement with the available experimental data. This is an essential improvement of the present calculation compared to the previous one \cite{Baklanov:10} in light of importance of verifying the values of \Eeff, $W_{T,P}$ and $W_{P}$, which is of primary interest here.

\section*{Acknowledgement}

The molecular calculations were partly performed on the Supercomputer ``Lomonosov''.
We acknowledge support from Saint Petersburg State University, research grant No.~0.38.652.2013, and RFBR, grant No.~13-02-01406. L.S.\ is also grateful to the President of Russian Federation grant No.~MK-5877.2014.2.

\bibliographystyle{./bib/apsrev}

\end{document}